# Are stealth scalar fields stable?

Valerio Faraoni[*] and Andres F. Zambrano Moreno[†]
*Physics Department, Bishop's University*
*2600 College Street, Sherbrooke,*
*Québec, Canada J1M 1Z7*

Non-gravitating (stealth) scalar fields associated with Minkowski space in scalar-tensor gravity are examined. Analytical solutions for both non-minimally coupled scalar field theory and for Brans-Dicke gravity are studied and their stability with respect to tensor perturbations is assessed using a covariant and gauge-invariant formalism developed for alternative gravity. For Brans-Dicke solutions, the stability with respect to homogeneous perturbations is also studied. There are regions of parameter space corresponding to stability and other regions corresponding to instability.

PACS numbers: 04.50.+h, 04.90.+e

## INTRODUCTION

It is a standard tenet of General Relativity (GR) that matter, energy, and stresses curve spacetime causing the Riemann tensor to be non-vanishing. The flat Minkowski space of Special Relativity corresponds to the absence of gravity, but when matter configurations described by the energy-momentum tensor $T_{\mu\nu}$ are introduced in the context of GR, spacetime becomes curved, as described by the Einstein equations

$$R_{\mu\nu} - \frac{1}{2} g_{\mu\nu} R = \kappa T_{\mu\nu} \tag{1}$$

(where $\kappa \equiv 8\pi G$, $G$ is Newton's constants, $R_{\mu\nu}$ is the Ricci tensor of the spacetime metric $g_{\mu\nu}$, and $R \equiv g^{\mu\nu} R_{\mu\nu}$ - we follow the notations of [1]).

The situation is somewhat different in theories of gravity alternative to GR. Here we focus on various scalar-tensor theories described by the general action in the Jordan frame [2, 3]

$$S_{ST} = \int d^4x \sqrt{-g} \left[ \psi(\phi) R - \frac{\omega(\phi)}{2} \nabla^\mu\phi \nabla_\mu \phi - V(\phi) \right]$$

$$+ S^{(m)}, \tag{2}$$

where $S^{(m)} = \int d^4x \sqrt{-g}\, \mathcal{L}^{(m)}$ is the matter action and $\psi$ and $\omega$ are coupling functions of the Brans-Dicke-like gravitational scalar field $\phi$ (a cosmological constant, if present, is incorporated in the scalar field potential $V(\phi)$). In special cases one can find Minkowski solutions resulting from the balance between matter and the Brans-Dicke-like scalar field $\phi$ or, in vacuo, between different parts of the effective energy-momentum tensor of $\phi$. When the scalar field $\phi$ is constant, the theory effectively turns into GR (with a cosmological constant if $V(\phi) \neq 0$) and solutions with constant $\phi$ are, therefore, trivial [35]. More interesting are Minkowski solutions with time-dependent scalar fields obtained as degenerate cases of Friedmann-Lemaître-Robertson-Walker (FLRW) classical solutions of Brans-Dicke gravity which we analyze in Sec. III. Analogs of these Minkowski spaces are known in string cosmology [5, 6].

In all these examples, the scalar field is non-trivial but recently, even more interesting solutions were found in which an *inhomogeneous*, wave-like field does not gravitate [7–13].

## NON-MINIMALLY COUPLED STEALTH FIELDS

Ayón-Beato *et al.* [7] found solutions for non-minimally coupled (*i.e.*, $\xi \neq 0$) scalar fields in $D$ spacetime dimensions in the scalar-tensor theory

$$S_{NMC} = \int d^D x \sqrt{-g} \left[ \frac{1}{2} \left( \frac{1}{\kappa} - \xi \phi^2 \right) R - \frac{1}{2} \nabla^\mu \phi \nabla_\mu \phi \right.$$

$$\left. - V(\phi) \right] \tag{3}$$

where

$$V(\phi) = \frac{2\xi^2}{(1-4\xi)^2} \left[ \lambda_1 \phi^{\frac{(1-2\xi)}{\xi}} + 8(D-1)(\xi - \xi_D)\lambda_2 \phi^{\frac{1}{2\xi}} \right] \tag{4}$$

if $\xi \neq 1/4$ and

$$V(\phi) = \frac{\lambda_2}{2} \phi^2 \left[ 2\ln\phi + \frac{\lambda_1}{\lambda_2} + D - 1 \right] \tag{5}$$

if $\xi = 1/4$ [7]. Here $\xi_D \equiv \frac{D-2}{4(D-1)}$ is the coupling constant corresponding to conformal invariance, and $\lambda_{1,2}$ are parameters.

The fields equations obtained from the variation of the action (2) can be rewritten in the form of effective Einstein equations $G_{\mu\nu} = 8\pi \Theta_{\mu\nu}$ and, by imposing that the effective stress-energy tensor of $\phi$ [14]

$$\Theta_{\mu\nu} \equiv \nabla_\mu \phi \nabla_\nu \phi - \frac{1}{2} g_{\mu\nu} \nabla^\alpha \phi \nabla_\alpha \phi - V g_{\mu\nu}$$

$$+ \xi \left( g_{\mu\nu} \Box - \nabla_\mu \nabla_\nu \right) \left( \phi^2 \right) \tag{6}$$

vanishes, the Minkowski metric $\eta_{\mu\nu}$ is a solution with non-trivial scalar $\phi$. Contrary to the stress-energy tensor of a minimally coupled scalar field in GR, the component $\Theta_{00}$, which is the energy density according to an observer at rest in Minkoswki space, is not positive-definite because of the second derivative terms on the right hand side of eq. (6). These non-canonical terms linear in the second derivatives of $\phi$ make the Minkowski solution possible: the canonical terms of $\Theta_{\mu\nu}$ quadratic in the first order derivatives of $\phi$ can be cancelled by the non-canonical terms containing second derivatives, which are responsible for the well known fact that non-minimally coupled scalar fields can violate all of the energy conditions [15].

The various solutions proposed by Ayón-Beato et al. as the parameters $(\xi, \lambda_1, \lambda_2)$ vary are summarized in the following (we assume $\xi \neq 0$ in the following because non-gravitating solutions are impossible for $\xi = 0$).

- $\xi \neq \xi_D, \frac{1}{4}; \quad \lambda_2 \neq 0, \lambda_1 < 0$:

$$\phi = \left( \frac{\lambda_2}{2} x^\mu x_\mu - \frac{\lambda_1}{2\lambda_2} \right)^{-\frac{2\xi}{1-4\xi}} \quad (7)$$

assuming $\lambda_1 < 0$ for regularity [7].

- $\xi \neq \xi_D, \frac{1}{4}; \quad \lambda_2 = 0$:

$$\phi = (k_\mu x^\mu)^{-\frac{2\xi}{1-4\xi}}, \quad k^\mu k_\mu = -|\lambda_1|. \quad (8)$$

- $\xi = \xi_D, \lambda_1 < 0, \alpha \equiv \lambda_2 \neq 0$:

$$\phi = \left( \frac{\alpha}{2} x^\mu x_\mu - \frac{\lambda_1}{2\alpha} \right)^{-\frac{(D-2)}{2}} \quad (9)$$

assuming $D > 2$ and $\lambda_1 < 0$ for regularity.

- $\xi = \frac{1}{4}, \lambda_2 \neq 0$:

$$\phi = \exp\left( \frac{\lambda_2}{2} x^\mu x_\mu - \frac{\lambda_1}{2\lambda_2} \right), \quad (10)$$

with

$$V(\phi) = \frac{\lambda_2}{2} \phi^2 \left( 2\ln\phi + \frac{\lambda_1}{\lambda_2} + D - 1 \right) \quad (11)$$

- $\xi = \frac{1}{4}, \lambda_2 = 0, \lambda_1 > 0$:

$$\phi = \exp(k_\mu x^\mu), \quad k^\mu k_\mu = \lambda_1 \quad (12)$$

with $V(\phi) = \frac{\lambda_1}{2}\phi^2$ (this solution is tachyonic [7]).

These *non-gravitating* or *stealth* solutions cannot be detected gravitationally [36]. Similarly, a stealth scalar field hovering above a BTZ black hole in 2+1 dimensions and non-gravitating, was found in [9]. Non-gravitating forms of matter defy intuition from GR and may be perceived as curiosities. However, their behaviour is exactly what would be needed to cure the cosmological constant problem, widely regarded as the most urgent problem of theoretical physics [16]: the energy density of quantum vacuum predicted with a straightforward, back-of-the-envelope calculation using extremely well established quantum mechanics is approximately 120 orders of magnitude larger than the measured cosmological energy density. It is as if the cosmological constant, while being present, does not gravitate. Therefore, it is interesting to study forms of non-gravitating matter with the hope of learning something useful for the cosmological constant problem. Moreover, non-gravitating scalar fields are impossible in GR and are found instead when scalar fields couple non-minimally to gravity, which is an unavoidable feature of string theories and higher order gravity theories inspired by attempts to renormalize gravity (it suffices to think of the dilaton of string theory [17, 18], or of the scalar-tensor representation of $f(R)$ gravity [19, 20].

It seems intuitive that the non-gravitating scalars in scalar-tensor theories are not "natural" solutions: one has to pick a special potential $V(\phi)$ and tune the model parameters in specific ranges in order to produce these solutions. As a consequence, one would suspect that these solutions may be unstable with respect to small perturbations and, therefore, unphysical for any practical purpose. Finding stable stealth solutions would increase their interest.

Unfortunately, the analysis of perturbations of these solutions is not easy due to the fact that they suffer from the well-known gauge-dependence problems associated with inhomogeneous perturbations. In this paper we study the stability of stealth solutions using the Bardeen-Ellis-Bruni-Hwang covariant and gauge-invariant formalism developed for the analysis of cosmological perturbations in alternative theories of gravity [21–23]. We make use of the fact that the Minkowski space hosting stealth scalars is a degenerate case of the spatially flat FLRW metric (but the inhomogeneous scalar $\phi$ is non-trivial).

A general vacuum action for a mixed scalar-tensor/$f(R)$ gravity in the metric formalism is (in the following we set $D = 4$)

$$S = \int d^4x \sqrt{-g} \left[ \frac{f(\phi, R)}{2} - \frac{\omega(\phi)}{2} \nabla^\mu \phi \nabla_\mu \phi - V(\phi) \right]. \quad (13)$$

For this theory, the field equations for the spatially flat FLRW universe with line element

$$ds^2 = -dt^2 + a^2(t)\left(dx^2 + dy^2 + dz^2\right) \quad (14)$$

are

$$H^2 = \frac{1}{3F} \left( \frac{\omega}{2}\dot\phi^2 + \frac{RF}{2} - \frac{f}{2} + V - 3H\dot F \right), \quad (15)$$

$$\dot H = -\frac{1}{2F}\left(\omega\dot\phi^2 + \ddot F - H\dot F\right), \quad (16)$$

$$\ddot\phi + 3H\dot\phi + \frac{1}{2\omega}\left(\frac{d\omega}{d\phi}\dot\phi^2 - \frac{\partial f}{\partial \phi} + 2\frac{dV}{d\phi}\right) = 0, \quad (17)$$

where $H \equiv \dot{a}/a$, $F \equiv \partial f/\partial R$, and an overdot denotes differentiation with respect to the comoving time $t$. In the Bardeen-Ellis-Bruni-Hwang-Vishniac formalism [21–23] the metric perturbations $A$, $B$, $H_L$, and $H_T$ are defined by

$$g_{00} = -a^2 (1 + 2AY) , \qquad (18)$$

$$g_{0i} = -a^2 BY_i , \qquad (19)$$

$$g_{ij} = a^2 [h_{ij}(1 + 2H_L) + 2H_T Y_{ij}] , \qquad (20)$$

where the scalar harmonics $Y$ are the eigenfunctions of the eigenvalue problem $\bar{\nabla}_i \bar{\nabla}^i Y = -k^2 Y$, $h_{ij}$ is the three-dimensional metric of the FLRW background space, $\bar{\nabla}_i$ is the covariant derivative operator of $h_{ij}$, and $k$ is an eigenvalue. The vector and tensor harmonics $Y_i$ and $Y_{ij}$ are given by

$$Y_i = -\frac{1}{k} \bar{\nabla}_i Y , \qquad (21)$$

and

$$Y_{ij} = \frac{1}{k^2} \bar{\nabla}_i \bar{\nabla}_j Y + \frac{1}{3} Y h_{ij} , \qquad (22)$$

respectively. The Bardeen gauge-invariant potentials [21] are

$$\Phi_H = H_L + \frac{H_T}{3} + \frac{\dot{a}}{k}\left(B - \frac{a}{k}\dot{H}_T\right) , \qquad (23)$$

$$\Phi_A = A + \frac{\dot{a}}{k}\left(B - \frac{a}{k}\dot{H}_T\right) + \frac{a}{k}\left[\dot{B} - \frac{1}{k}\left(a\dot{H}_T\right)^{\cdot}\right] \qquad (24)$$

and the Ellis-Bruni [22] variable is

$$\triangle \phi = \delta \phi + \frac{a}{k} \dot{\phi}\left(B - \frac{a}{k}\dot{H}_T\right) , \qquad (25)$$

with similar relations defining the gauge-invariant variables $\triangle f$, $\triangle F$, and $\triangle R$. To first order, the gauge-invariant perturbations satisfy the (redundant) system [23]

$$\triangle \ddot{\phi} + \left(3H + \frac{\dot{\phi}}{\omega}\frac{d\omega}{d\phi}\right)\triangle \dot{\phi} + \left[\frac{k^2}{a^2} + \frac{\dot{\phi}^2}{2}\frac{d}{d\phi}\left(\frac{1}{\omega}\frac{d\omega}{d\phi}\right) - \frac{d}{d\phi}\left(\frac{1}{2\omega}\frac{\partial f}{\partial \phi} - \frac{1}{\omega}\frac{dV}{d\phi}\right)\right]\triangle \phi$$
$$= \dot{\phi}\left(\dot{\Phi}_A - 3\dot{\Phi}_H\right) + \frac{\Phi_A}{\omega}\left(\frac{\partial f}{\partial \phi} - 2\frac{dV}{d\phi}\right) + \frac{1}{2\omega}\frac{\partial^2 f}{\partial \phi \partial R}\triangle R , \qquad (26)$$

$$\triangle \ddot{F} + 3H\triangle \dot{F} + \left(\frac{k^2}{a^2} - \frac{R}{3}\right)\triangle F + \frac{F}{3}\triangle R + \frac{2}{3}\omega\dot{\phi}\triangle \dot{\phi} + \frac{1}{3}\left(\dot{\phi}^2\frac{d\omega}{d\phi} + 2\frac{\partial f}{\partial \phi} - 4\frac{dV}{d\phi}\right)\triangle \phi$$
$$= \dot{F}\left(\dot{\Phi}_A - 3\dot{\Phi}_H\right) + \frac{2}{3}(FR - 2f + 4V)\Phi_A , \qquad (27)$$

$$\ddot{H}_T + \left(3H + \frac{\dot{F}}{F}\right)\dot{H}_T + \frac{k^2}{a^2}H_T = 0 , \qquad (28)$$

$$-\dot{\Phi}_H + \left(H + \frac{\dot{F}}{2F}\right)\Phi_A = \frac{1}{2}\left(\frac{\triangle\dot{F}}{F} - H\frac{\triangle F}{F} + \frac{\omega}{F}\dot{\phi}\triangle\phi\right) , \qquad (29)$$

$$\left(\frac{k}{a}\right)^2 \Phi_H + \frac{1}{2}\left(\frac{\omega}{F}\dot{\phi}^2 + \frac{3}{2}\frac{\dot{F}^2}{F^2}\right)\Phi_A = \frac{1}{2}\left\{\frac{3}{2}\frac{\dot{F}\triangle\dot{F}}{F^2} + \left(3\dot{H} - \frac{k^2}{a^2} - \frac{3H}{2}\frac{\dot{F}}{F}\right)\frac{\triangle F}{F}\right.$$
$$\left. + \frac{\omega}{F}\dot{\phi}\triangle\dot{\phi} + \frac{1}{2F}\left[\dot{\phi}^2\frac{d\omega}{d\phi} - \frac{\partial f}{\partial \phi} + 2\frac{dV}{d\phi} + 6\omega\dot{\phi}\left(H + \frac{\dot{F}}{2F}\right)\right]\triangle\phi\right\} , \qquad (30)$$

$$\Phi_A + \Phi_H = -\frac{\triangle F}{F} , \qquad (31)$$

$$\ddot{\Phi}_H + H\dot{\Phi}_H + \left(H + \frac{\dot{F}}{2F}\right)\left(2\dot{\Phi}_H - \dot{\Phi}_A\right) + \frac{1}{2F}\left(f - 2V - RF\right)\Phi_A$$
$$= -\frac{1}{2}\left[\frac{\Delta \ddot{F}}{F} + 2H\frac{\Delta \dot{F}}{F} + (P - \rho)\frac{\Delta F}{2F} + \frac{\omega}{F}\dot{\phi}\Delta\dot{\phi} + \frac{1}{2F}\left(\dot{\phi}^2 \frac{d\omega}{d\phi} + \frac{\partial f}{\partial \phi} - 2\frac{dV}{d\phi}\right)\Delta\phi\right], \tag{32}$$

and

$$\Delta R = 6\left[\ddot{\Phi}_H + 4H\dot{\Phi}_H + \frac{2}{3}\frac{k^2}{a^2}\Phi_H - H\dot{\Phi}_A - \left(2\dot{H} + 4H^2 - \frac{k^2}{3a^2}\right)\Phi_A\right]. \tag{33}$$

These equations simplify considerably in a Minkowski background. However, due to the fact that the background stealth scalar field is inhomogeneous and time-dependent, solving the simplified equations for the coupled gauge-invariant variables still constitutes a highly non-trivial task. Fortunately, eq. (28) for the tensor mode $H_T$ decouples, to first order, from the remaining equations. In the zero momentum limit $k \to 0$ one obtains the first integral

$$\dot{H}_T = \frac{C}{\psi(\phi)} \tag{34}$$

where $C$ is an integration constant and we have set $f(\phi, R) \equiv \psi(\phi) R$ for scalar-tensor gravity. The non-minimally coupled theory studied by [7] is, for all purposes, a scalar-tensor gravity with $\psi(\phi) = \frac{1}{\kappa} - \xi\phi^2$ and $\omega = 1$. The first integral of motion (34) allows one to draw conclusions about the time evolution of the tensor perturbation $H_T$ in this theory.

For $\xi \neq \xi_D, \frac{1}{4}$, $\lambda_2 \neq 0$, and $\lambda_1 < 0$, the perturbation of the solutions given by $\eta_{\mu\nu}$ and eq. (7) obeys

$$\dot{H}_T = \frac{\kappa C}{1 - \xi\kappa\left[\frac{\lambda_2}{2}x^\mu x_\mu + \frac{|\lambda_1|}{2\lambda_2}\right]^{-\frac{4\xi}{(1-4\xi)}}} \tag{35}$$

where $-4\xi/(1 - 4\xi)$ is positive for $\xi < 0$ and for $\xi > 1/4$ and negative otherwise, and $\lambda_2 x^\mu x_\mu = -\lambda_2 t^2 + \lambda_2 r^2$ where $r \equiv \sqrt{x^2 + y^2 + z^2}$. Therefore, if $\xi < 0$ or $\xi > 1/4$, $\dot{H}_T \to 0$ as $t \to +\infty$ for any value of the constant $C$.

If $0 < \xi < 1/4$, $\dot{H}_T \to \kappa C$ and if $C > 0$, the gauge-invariant variable $H_T$ grows linearly with time showing an instability of the Ayón-Beato et al. solution, while it dies off if $C \leq 0$. Since $C$ is determined by the initial conditions on $\dot{H}_T$, which must be arbitrary, we conclude that this case is unstable.

For $\xi \neq \xi_D, \frac{1}{4}$ and $\lambda_2 = 0$, the relevant solution is $\eta_{\mu\nu}$ with eq. (8) and

$$\dot{H}_T = \frac{\kappa C}{1 - \kappa\xi\left(k_\mu x^\mu\right)^{-\frac{4\xi}{1-4\xi}}}. \tag{36}$$

If $\xi < 0$ or $\xi > 1/4$, $\dot{H}_T \to 0$ as $t \to +\infty$ corresponds again to stability. If $0 < \xi < 1/4$, then $\dot{H}_T \to \kappa C$, and we obtain the same conclusions as in the previous cases.

For $\xi = \xi_D = 1/6$ in $D = 4$, $\lambda_1 < 0$ and $\alpha \neq 0$, the solution is given by eq. (9) and

$$\dot{H}_T = \frac{\kappa C}{1 - \frac{\kappa\xi}{\left(\frac{\alpha}{2}x^\mu x_\mu - \frac{\lambda_1}{2\alpha}\right)^2}} \tag{37}$$

and $\dot{H}_T \to \kappa C$ with instability if the initial conditions on the perturbations are such that $C > 0$. Again, we conclude that this case is unstable.

For $\xi = \frac{1}{4}$ and $\lambda_2 \neq 0$, the relevant solution is given by eq. (10) and

$$\dot{H}_T = \frac{\kappa C}{1 - \kappa\xi\exp\left(\lambda_2 x^\mu x_\mu - \frac{\lambda_1}{\lambda_2}\right)}. \tag{38}$$

If the parameter $\lambda_2$ is positive, then $\dot{H}_T \to \kappa C$ as $t \to +\infty$ and there is an instability if $C > 0$ while, if $\lambda_2 < 0$ then $\dot{H}_T \to 0$ for any value of $C$ and the solution is stable irrespective of the initial conditions which determine the value of $C$.

For $\xi = 1/4$, $\lambda_2 = 0$ and $\lambda_1 > 0$, the solution given by eq. (12) yields

$$\dot{H}_T = \frac{\kappa C}{1 - \xi\exp\left(2k_\mu x^\mu\right)} \tag{39}$$

and, again, $\dot{H}_T \to \kappa C$ with an instability if $C > 0$; since the initial conditions must be arbitrary, this case is also unstable. The conditions for the stability of non-minimally coupled stealth fields are summarized in Table I.

### NON-GRAVITATING BRANS-DICKE SOLUTIONS WITH NON-CONSTANT SCALAR FIELD

Another class of solutions of scalar-tensor theories exhibiting the gravitational Cheshire effect is known in

Brans-Dicke gravity. These solutions include degenerate cases of classical solutions of Brans-Dicke cosmology and also recent solutions found by Robinson [8]. The action is given by eq. (13) with $f(\phi, R) = \phi R$ and $\omega =$const.

### A non-gravitating Nariai solution

The Nariai solution of Brans-Dicke theory [24–26] corresponds to a perfect fluid with constant equation of state $P = (\gamma - 1)\rho$ (with $\gamma =$constant) and is given by the spatially flat FLRW metric (14) with $V(\phi) \equiv V_0 =$const.,

$$a(t) = a_0 (1 + \lambda t)^q, \quad (40)$$

$$\phi_0(t) = \phi_* (1 + \lambda t)^s, \quad (41)$$

$$q = \frac{2[\omega(2-\gamma) + 1]}{3\omega\gamma(2-\gamma) + 4}, \quad (42)$$

$$s = \frac{2(4 - 3\gamma)}{3\omega\gamma(2-\gamma) + 4}, \quad (43)$$

$$\rho = \rho_0 (1 + \lambda t)^{-3\gamma q} \quad (44)$$

where $a_0, \lambda, \phi_*$, and $\rho_0$ are constants and $s + 3\gamma q = 2$. The special case $\gamma = 0$ corresponds to the cosmological constant (treated as a perfect fluid) and shows that the natural cosmological solution of Brans-Dicke gravity with only a cosmological constant is not de Sitter space, but a power-law expanding universe. Historically, this feature was the foundation of the extended and hyperextended inflationary scenarios ([27], see also [28]). For $\gamma = 0$ and $\omega = -1/2$ one obtains Minkowski space with

$$a = \text{const.}, \quad \phi_0(t) = \phi_* (1 + \lambda t)^2, \quad (45)$$

and $\lambda = \sqrt{\frac{8\pi V_0}{\phi_*^2}} > 0$. A stealth Brans-Dicke field cancels the cosmological constant and provides flat spacetime. Applying eq. (34) to this case, we have

$$\dot{H}_T = \frac{C}{\phi_*(1 + \lambda t)^2}, \quad (46)$$

which vanishes asymptotically as $t \to +\infty$ for any value of the integration constant $C$, leading to stability of this Minkowski space with respect to tensor perturbations.

In this case, since the unperturbed scalar field $\phi_0(t)$ is homogeneous, it is meaningful to consider also *homogeneous* perturbations of this Minkowski space. By assuming that

$$H(t) = \delta H(t), \quad \phi(t) = \phi_0(t) + \delta\phi(t) \quad (47)$$

and using the Brans-Dicke field equations for a spatially flat FLRW metric

$$\dot{H} = -\frac{\omega}{2}\left(\frac{\dot\phi}{\phi}\right)^2 + 2H\frac{\dot\phi}{\phi} + \frac{1}{2(2\omega+3)\phi}\left(\phi\frac{dV}{d\phi} - 2V\right), \quad (48)$$

$$\ddot\phi + 3H\dot\phi = \frac{1}{2\omega + 3}\left(-\phi\frac{dV}{d\phi} + 2V\right), \quad (49)$$

one obtains the first order evolution equations for the homogeneous perturbations

$$\delta\dot H = \left[-\frac{1}{2}\left(\frac{\dot\phi_0}{\phi_0}\right)^2 + \frac{V_0}{2\phi_0^2}\right]\delta\phi + \frac{1}{2}\frac{\dot\phi_0}{\phi_0}\delta\dot\phi + 2\frac{\dot\phi_0}{\phi_0}\delta H, \quad (50)$$

$$\delta\ddot\phi + 3\dot\phi_0 \delta H = 0. \quad (51)$$

The use of eqs. (50) and (45) in eq. (51) yields

$$\delta\dddot\phi - \left(\frac{5\lambda}{1+\lambda t}\right)\delta\ddot\phi + \left(\frac{6\lambda^2}{(1+\lambda t)^2}\right)\delta\dot\phi$$
$$+ \frac{3\lambda(V_0 - 4\lambda^2\phi_*)}{\phi_*(1+\lambda t)^3}\delta\phi = 0. \quad (52)$$

The power-law ansatz

$$\delta\phi(t) = \delta_0 (1 + \lambda t)^s \quad (53)$$

(with $\delta_0$ and $s$ constants) yields the algebraic cubic equation

$$\varphi(s) \equiv s^3 - 8s^2 + 13s + 3\left(\frac{V_0}{\lambda^2 \phi_*} - 4\right) = 0. \quad (54)$$

Remembering that the roots of the cubic equation $ax^3 + bx^2 + cx + d = 0$ are given by

$$r_1 = \alpha^{1/3} - \beta^{1/3}, \quad (55)$$

$$r_2 = \alpha^{\frac{1}{3}} e^{\frac{2\pi i}{3}} - \beta^{\frac{1}{3}} e^{\frac{4\pi i}{3}}, \quad (56)$$

$$r_3 = \alpha^{\frac{1}{3}} e^{\frac{4\pi i}{3}} - \beta^{\frac{1}{3}} e^{\frac{2\pi i}{3}}, \quad (57)$$

where

$$\alpha \equiv \frac{-q + \sqrt{\triangle}}{2}, \quad \beta \equiv \frac{-q - \sqrt{\triangle}}{2}, \quad (58)$$

$$\triangle = 4p^3 + q^2 \quad (59)$$

is the discriminant, and

$$p \equiv \frac{3ac - b^2}{9a^2}, \quad (60)$$

$$q \equiv \frac{2b^3 - 9abc + 27a^2 d}{27a^3}, \quad (61)$$

it is easily seen that

$$\triangle = \frac{1}{27^2} \left\{ 62500 + \left[ -88 + 81 \left( \frac{V_0}{\phi_* \lambda^2} - 4 \right) \right]^2 \right\} \quad (62)$$

is positive. This fact implies that eq. (54) admits only one real root and two complex conjugate ones. The real root

$$r_1 = \frac{1}{3 \cdot 2^{1/3}} \left\{ \left[ 88 - 81 \left( \frac{V_0}{\phi_* \lambda^2} - 4 \right) \right] + \sqrt{62500 + \left[ -88 + 81 \left( \frac{V_0}{\phi_* \lambda^2} - 4 \right) \right]^2} \right\}^{1/3}$$

$$- \frac{1}{3 \cdot 2^{1/3}} \left\{ \left[ 88 - 81 \left( \frac{V_0}{\phi_* \lambda^2} - 4 \right) \right] - \sqrt{62500 + \left[ -88 + 81 \left( \frac{V_0}{\phi_* \lambda^2} - 4 \right) \right]^2} \right\}^{1/3} \quad (63)$$

is positive, hence the mode $\delta\phi = \delta_0 (1 + \lambda t)^{r_1}$ grows without bound as $t \to +\infty$. Technically speaking, the perturbed solution $(H, \phi)$ "runs away" from the unperturbed space $(0, \phi_0)$, but one should ask instead whether the perturbations destroy the Minkowski space, or whether the latter remains Minkowskian. To answer this question, consider

$$\delta H = -\frac{\delta\ddot{\phi}}{3\dot{\phi}} = -\frac{r_1 (r_1 - 1) \lambda^2}{6\lambda\phi_*} \delta_0 (1 + \lambda t)^{r_1 - 3} . \quad (64)$$

The perturbed Hubble parameter $\delta H$ decays if $r_1 < 3$ and stays constant if $r_1 > 3$. The polynomial

$$\varphi(s) \equiv s^3 - 8s^2 + 13s + 3 \left( \frac{V_0}{\lambda^2 \phi_*} - 4 \right) \quad (65)$$

crosses the $s$-axis only at $s = r_1$ and, since $\varphi(3) = 3\left(\frac{V_0}{\lambda_2 \phi_*} - 6\right)$, it is $r_1 < 3$ when $\varphi(3) > 0$, or $V_0 > 6\lambda^2 \phi_*$; $r_1 = 3$ for $V = 6\lambda^2 \phi_*$, and $r_1 > 3$ when $V_0 < 6\lambda^2 \phi_*$. Therefore, the mode $\delta\phi_1 = \delta_0 (1 + \lambda t)^{r_1}$ is *stable* (in the sense that the solution remains non-gravitating) for $V_0 \geq 6\lambda^2 \phi_*$ and *unstable* otherwise.

We still need to assess the stability of the other two modes corresponding to the roots $r_{2,3}$ of eq. (54).

The real part of the roots $r_{2,3}$, which determines the growing or decaying behaviour of $(1 + \lambda t)^{r_{2,3} - 3}$, is

$$\mathcal{R}e(r_2) = \mathcal{R}e(r_3) = \frac{1}{2} \left( \beta^{1/3} - \alpha^{1/3} \right) = -\frac{r_1}{2} \quad (66)$$

and we need to assess whether $\mathcal{R}e(r_2) = \mathcal{R}e(r_3)$ is less than or equal to 3 in the region of parameter space in which the mode $\delta\phi_1$ is stable. Note that $\varphi(-6) > 0$, which corresponds to $r_1 < -6$ and $\mathcal{R}e(r_2) = \mathcal{R}e(r_3) > 3$ corresponds to $V_0 > 198\lambda^2 \phi_*$, hence the modes $\delta\phi_{2,3} = \delta_0 (1 + \lambda t)^{r_{2,3}}$ are *unstable* for this range of parameters, and *stable* otherwise.

Putting together the previous considerations for all the independent modes $\delta\phi_{1,2,3}$, one obtains that the solution $(H, \phi) = (0, \phi_0)$ is linearly stable (in the sense that the solution remains non-gravitating) for

$$6\lambda^2 \phi_* \leq V_0 \leq 198\lambda^2 \phi_* ; \quad (67)$$

outside of this parameter range perturbations grow without bound destroying Minkowski space.

### Minkowski spaces with exponential scalar fields

The phase plane $(H, \phi)$ of spatially flat FLRW cosmology with [37] $V(\phi) = \Lambda\phi$ was studied in [32]. Two de Sitter fixed points are always present in the phase plane:

$$a_{(\pm)}(t) = a_0 \exp\left[ \pm (\omega + 1) \sqrt{\frac{2\Lambda}{(2\omega + 3)(3\omega + 4)}} \, t \right] , \quad (68)$$

$$\phi_{(\pm)}(t) = \phi_0 \exp\left[ \pm \sqrt{\frac{2\Lambda}{(2\omega + 3)(3\omega + 4)}} \, t \right] . \quad (69)$$

These solutions were found in [31–33]. For $\omega = -1$ one obtains [38] Minkowski space with an exponentially expanding/contracting scalar field

$$a_{(\pm)} \equiv 1 , \quad \phi_{(\pm)}(t) = \phi_0 \exp\left( \pm\sqrt{2\Lambda} \, t \right) . \quad (70)$$

Eq. (34) yields

$$\dot{H}_T = \frac{C}{\phi_0} \exp\left[ \mp \sqrt{2\Lambda} \, t \right] . \quad (71)$$

Since $\phi_0 > 0$ in order to keep the gravitational coupling positive, for $\phi_+(t)$, $\dot{H}_T$ tends to zero as $t \to +\infty$ for any choice of initial conditions corresponding to *stability*. For $\phi_-(t)$, instead, $\dot{H}_T$ diverges, corresponding to *instability*.

One can give a physical interpretation of this result: the effective gravitational coupling $G_{eff} \sim 1/\phi$ decreases in the first case and increases in the second one. If $G_{eff}$ increases and diverges with time, any gravitational perturbation of Minkowski space will become stronger, making the deviation from flatness more pronounced, in a positive feedback mechanism. If instead $G_{eff}$ tends to zero as time progresses, the effect of perturbations from flatness become less and less pronounced, contributing to *stability*.

Again, it is easy to study *homogeneous* perturbations of the solution (70): eqs. (47)-(49) now yield the evolution equations for the perturbations $\left(\delta H_{(\pm)}, \delta\phi_{(\pm)}\right)$

$$\delta \dot{H}_{(\pm)} = -\frac{\left(\dot{\phi}_0^{(\pm)}\right)^2}{\left(\phi_0^{(\pm)}\right)^3} \delta\phi_{(\pm)} + \frac{\dot{\phi}_0^{(\pm)}}{\left(\phi_0^{(\pm)}\right)^2} \delta\dot{\phi}_{(\pm)}$$

$$+2\frac{\dot{\phi}_0^{(\pm)}}{\phi_0^{(\pm)}} \delta H_{(\pm)}, \qquad (72)$$

$$\delta\dddot{\phi}_{(\pm)} + 3\dot{\phi}_0^{(\pm)} \delta H_{(\pm)} = \Lambda\delta\phi_{(\pm)}. \qquad (73)$$

By using eq. (73) in eq. (72), one obtains

$$\delta\dddot{\phi}_{(\pm)} - \left(\frac{\ddot{\phi}_0^{(\pm)}}{\dot{\phi}_0^{(\pm)}} + 2\frac{\dot{\phi}_0^{(\pm)}}{\phi_0^{(\pm)}}\right) \delta\ddot{\phi}_{(\pm)}$$

$$+ \left[3\left(\frac{\dot{\phi}_0^{(\pm)}}{\phi_0^{(\pm)}}\right)^2 - \Lambda\right] \delta\dot{\phi}_{(\pm)}$$

$$+ \left[\Lambda\frac{\ddot{\phi}_0^{(\pm)}}{\dot{\phi}_0^{(\pm)}} - 3\left(\frac{\dot{\phi}_0^{(\pm)}}{\phi_0^{(\pm)}}\right)^3 + 2\Lambda\frac{\dot{\phi}_0^{(\pm)}}{\phi_0^{(\pm)}}\right] \delta\phi_{(\pm)} = 0, \qquad (74)$$

and the further use of eq. (70) yields

$$\delta\dddot{\phi}_{(\pm)} \mp 3\sqrt{2\Lambda}\,\delta\ddot{\phi}_{(\pm)} + 5\Lambda\delta\dot{\phi}_{(\pm)} \mp 3\Lambda\sqrt{2\Lambda}\,\delta\phi_{(\pm)} = 0. \qquad (75)$$

The associated algebraic equation

$$\varphi_{(\pm)}(s) \equiv s^3 \mp 3\sqrt{2\Lambda}\,s^2 + 5\Lambda s + \dot{\phi} \mp 3\Lambda\sqrt{2\Lambda} = 0 \qquad (76)$$

has discriminant

$$\triangle_{(\pm)} = 4p^3 + q^2 = \frac{212}{27}\Lambda^3 > 0 \qquad (77)$$

and therefore eq. (76) admits one real root $r_1$ and two complex conjugate roots $r_1, r_2$. Since

$$\alpha_{(\pm)} = \frac{-q + \sqrt{\triangle}}{2} = \Lambda^{3/2}\left(\pm\sqrt{2} + \sqrt{\frac{53}{27}}\right), \qquad (78)$$

$$\beta_{(\pm)} = \frac{-q - \sqrt{\triangle}}{2} = \Lambda^{3/2}\left(\pm\sqrt{2} - \sqrt{\frac{53}{27}}\right), \qquad (79)$$

the real root (for each of the upper/lower sign solutions) is

$$r_1^{(+)} = \alpha_{(+)}^{1/3} - \beta_{(+)}^{1/3} = \sqrt{\Lambda}\left[\left(\sqrt{2} + \sqrt{\frac{53}{27}}\right)^{1/3} + \left(\sqrt{\frac{53}{27}} - \sqrt{2}\right)^{1/3}\right], \qquad (80)$$

$$r_1^{(-)} = \alpha_{(-)}^{1/3} - \beta_{(-)}^{1/3} = \sqrt{\Lambda}\left[\left(\sqrt{\frac{53}{27}} - \sqrt{2}\right)^{1/3} + \left(\sqrt{\frac{53}{27}} + \sqrt{2}\right)^{1/3}\right]. \qquad (81)$$

Since $r_1^{(\pm)} > 0$, the corresponding mode $\delta\phi_1^{(\pm)} = \delta_0 \exp\left(r_1^{(\pm)}t\right)$ grows away from $\phi_0^{(\pm)}$ without bound. However, we want to know if the spacetime remains Minkowskian or not, which is obtained by considering

$$\delta H^{(\pm)} = \frac{\Lambda\delta\phi_{(\pm)} - \delta\ddot{\phi}_{(\pm)}}{3\dot{\phi}_{(\pm)}}$$

$$= \frac{\pm\left(\Lambda - r_{(\pm)}^2\right)}{3\phi_*\sqrt{2\Lambda}}\delta_0 \exp\left[\left(r_{(\pm)} \mp \sqrt{2\Lambda}\right)t\right]. \qquad (82)$$

We want to know whether $r_{(\pm)} \mp \sqrt{2\Lambda}$ (rather than $r_{(\pm)}$) is positive. It is easy to see that the inequality $r_1^{(+)} - \sqrt{2\Lambda} > 0$ is never satisfied, hence this mode is always *stable* (in the sense that the solution remains non-gravitating), while the inequality $r_1^{(-)} + \sqrt{2\Lambda} > 0$ can never be satisfied, and this mode is *unstable*, therefore the solution $(H, \phi) = \left(0, \phi_0^{(-)}\right)$ is *unstable*. It remains to check the other two perturbation modes for the $\left(0, \phi_0^{(+)}\right)$ solution when $\delta H_1^{(+)}$ is stable.

The complex conjugate roots $r_2, r_3$ of the algebraic equation $\varphi_{(+)}(s) = 0$ have real part

$$\mathcal{R}e\left(r_2^{(+)}\right) = \mathcal{R}e\left(r_3^{(+)}\right) = \frac{1}{2}\left(\beta^{1/3} - \alpha^{1/3}\right) = -\frac{r_1^{(+)}}{2} \tag{83}$$

and the inequality

$$\mathcal{R}e\left(r_{2,3}^{(+)}\right) - \sqrt{2\Lambda} > 0 \tag{84}$$

is never satisfied, hence also the perturbations $\left(\delta H_{2,3}^{(+)}, \delta\phi_{2,3}^{(+)}\right)$ do not gravitate and the space $\left(0, \phi_0^{(+)}\right)$ remains a non-gravitating solution of the Brans-Dicke field equations.

### Robinson's solutions of Brans-Dicke gravity

Robinson [8] has considered Minkowski space solutions of the coupled Brans-Dicke-Maxwell equations with non-trivial scalar field. Even simpler are the solutions without electromagnetic field, given by the metric $\eta_{\mu\nu}$ and

$$\phi_0(u) = \begin{cases} (c_1 u + c_2)^{1/(\omega_0+1)}, & \omega \neq -1 \\ \exp(c_1 u + c_2), & \omega = -1, \end{cases} \tag{85}$$

where $u = (t - x)/\sqrt{2}$ is the usual retarded coordinate in Minkowski space and $c_1, c_2$ are integration constants (which assume the role of parameters in the unperturbed solution). Applying again eq. (34) one obtains, for $\omega \neq -1$,

$$\dot{H}_T = \frac{C}{(c_1 u + c_2)^{\frac{1}{\omega+1}}}. \tag{86}$$

It is easy to see that, if $\omega > -1$, $\dot{H}_T \to 0$ as $t \to +\infty$ irrespective of the initial conditions (*i.e.*, of the value of $C$) and of the sign of $c_1$, corresponding to *stability*.

If instead $\omega < -1$, then $\dot{H}_T \to +\infty$ (unless $C = 0$) and there is *instability*.

For $\omega = -1$, tensor mode perturbations obey

$$\dot{H}_T = \frac{C}{\exp(c_1 u + c_2)}; \tag{87}$$

as $t \to +\infty$, also $u \to +\infty$ and $H_T$ grows if the initial conditions are such that $C > 0$, therefore this solution is *unstable*.

### CONCLUSIONS

Non-gravitating matter is interesting in principle because it could potentially teach us lessons of some relevance for the cosmological constant problem [16]. The linear stability of stealth solutions with respect to small tensor perturbations, analyzed in a covariant and gauge-invariant way is, therefore, physically interesting. The solution of the full system of differential equations for the gauge-invariant perturbations is, in general, a daunting task. However, the equation for the tensor mode $H_T$ is very simple and decouples from the other equations (to first order). While one would expect stealth solutions to be contrived and extremely unstable, instability with respect to tensor modes shows up only in certain regions of the parameter space, while there are regions corresponding to stability. Table I summarizes the situation for non-minimally coupled stealth scalar fields. Similarly, non-gravitating scalar fields in Brans-Dicke theory show regions of stability in their parameter spaces. For these solutions it was possible to analyze also the stability with respect to homogeneous perturbations.

Our results should be regarded as preliminary; one should still assess stability with respect to scalar perturbations, and stability to order higher than linear in order to draw more definitive conclusions. At the moment, there is hope to find *stable* stealth fields.

This work is supported by the Natural Sciences and Engineering Research Council of Canada (NSERC).

| $\xi$ | $\lambda_1$ | $\lambda_2$ | Stability |
|---|---|---|---|
| $\xi \neq \xi_D$, $\xi < 0$ or $\xi > \frac{1}{4}$ | any | $\lambda_2 \neq 0$ | stable |
| $\xi \neq \xi_D$, $0 < \xi < \frac{1}{4}$ | any | $\lambda_2 \neq 0$ | unstable |
| $\xi \neq \xi_D$, $\xi < 0$ or $\xi > \frac{1}{4}$ | any | $\lambda_2 = 0$ | stable |
| $\xi \neq \xi_D$, $0 < \xi < \frac{1}{4}$ | any | $\lambda_2 = 0$ | unstable |
| $\xi = \xi_D$ | $\lambda_1 < 0$ | any | unstable ($\alpha \neq 0$) |
| $\xi = \frac{1}{4}$ | any | $\lambda_2 > 0$ | unstable |
| $\xi = \frac{1}{4}$ | any | $\lambda_2 < 0$ | stable |
| $\xi = \frac{1}{4}$ | $\lambda_1 > 0$ unphysical | $\lambda_2 = 0$ | unstable |

TABLE I: A summary of the stability analysis of the Minkowski spacetime solutions with non-minimally coupled stealth scalar fields.